\documentstyle[11pt,aaspp4]{article}

\received{}
\accepted{}
\journalid{}{}
\articleid{}{}

\slugcomment{\it To appear in ApJ Letters}

\lefthead{Roth \& Bauer}
\righthead{C I Absorption Toward PKS 1756$+$237}

\begin{document}

\newcommand{\NHI}{$N({\rm H\,I})$}

\title{The $z =$ 1.6748 C I Absorber Toward PKS 1756$+$237}

\author{Katherine C. Roth\altaffilmark{1,2,3} and James M. Bauer\altaffilmark{2}}
\affil{Institute for Astronomy, University of Hawaii, 2680 Woodlawn Drive, Honolulu, HI 96825}

\altaffiltext{1}{Hubble Fellow}
\altaffiltext{2}{Visiting Astronomer, W.M. Keck Observatory, 
operated as a scientific partnership among the California Institute of
Technology, the University of California and NASA.  WMKO was made
possible by the generous financial support of the W.M. Keck
Foundation.}
\altaffiltext{3}{current address:  Dept.\ of Physics \& Astronomy, Johns
Hopkins University, 3400 N.\ Charles St., Baltimore, MD 21218 ---  e-mail: kroth@pha.jhu.edu}
 
\begin{abstract}
We report the detection of the $\lambda$1560 and $\lambda$1657
ground-state C I absorption features in the $z_{\rm abs}=1.6748$ system
toward the QSO PKS 1756$+$237.  We find no associated C I* lines with a
resulting 3$\,\sigma$ excitation temperature upper-limit of $T_{\rm
ex}\leq 8.54\,(+0.65,\,-0.56)$ K, which is consistent with the
predicted CMBR temperature of $T_{\rm CMBR}=7.291$ K.  Because the
redshifted CMBR populates the $J=1$ level and leaves little room for
additional local excitation through either collisions or UV pumping,
our data place 2--3 times more stringent limits on particle densities
and UV fields than existing {\it Copernicus}\/ observations of similar
column density sightlines in the Milky Way.  We also detect several Ni
II lines and the weak Fe II $\lambda$1611 line.  From the Ni/Fe column
density ratio we find evidence for dust at a dust-to-metals ratio
similar to that seen toward warm Galactic disk clouds.  Based on these
findings and supported by our Ly$\alpha$ spectrum we propose
to reclassify this system as a damped Ly$\alpha$ absorber.
\end{abstract}

\keywords{cosmic microwave background --- quasars: absorption lines --- quasars: individual (PKS 1756$+$237) --- ISM: abundances --- dust}

\section{Introduction}

Absorption spectra of QSOs sample the distribution and physical state
of intervening gaseous material in the early universe.  Damped
Ly$\alpha$ systems (log \NHI $\gtrsim$ 20.3 cm$^{-2}$) are particularly
important since they are believed to originate within young galaxies or
galaxy building blocks.  The simple ionization state and high column
densities within these systems enable the determination of accurate
abundances for many species.  Metallicities are commonly based on Zn II
and S II measurements, while Fe-peak elements, eg.\ Cr II, Fe II, and
Ni II, are used to infer depletion estimates of dust grain formation.
Abundances of $\alpha$-process elements, eg.\ Si II, Mn II, and Ti II,
may reflect the importance of high-mass stars to the enrichment of the
interstellar medium (ISM) within the absorbers.  The excitation of C I
fine-structure levels depends upon the gas pressure.  Also, redshifted
Cosmic Microwave Background Radiation (CMBR) photons excite C I more
efficiently than the present-day CMBR due to the predicted temperature
dependence on redshift, $T=T_{\rm CMBR} (1+z)$.  Therefore,
measurements of C I excitation in damped Ly$\alpha$ systems can yield
sensitive upper-limits on the temperature of the CMBR in the past,
testing the fundmental prediction of simple big bang cosmologies that
the universe cools as it expands.  Neutral species like C I are
associated in the Milky Way ISM primarily with dense, molecule-rich
lines of sight.  Only three high-redshift C I absorbers are known
(\cite{gej97}, \cite{son94}, \cite{mey86}, \cite{bla85}, \cite{bla82}),
consistent with the apparent lack of molecules in many damped
Ly$\alpha$ systems.  In this Letter we report the unexpected detection
of C I absorption at $z=1.6748$ toward the QSO PKS 1756$+$237.

\section{Observations and Analysis}

\begin{figure}
\epsscale{0.85}
\plotone{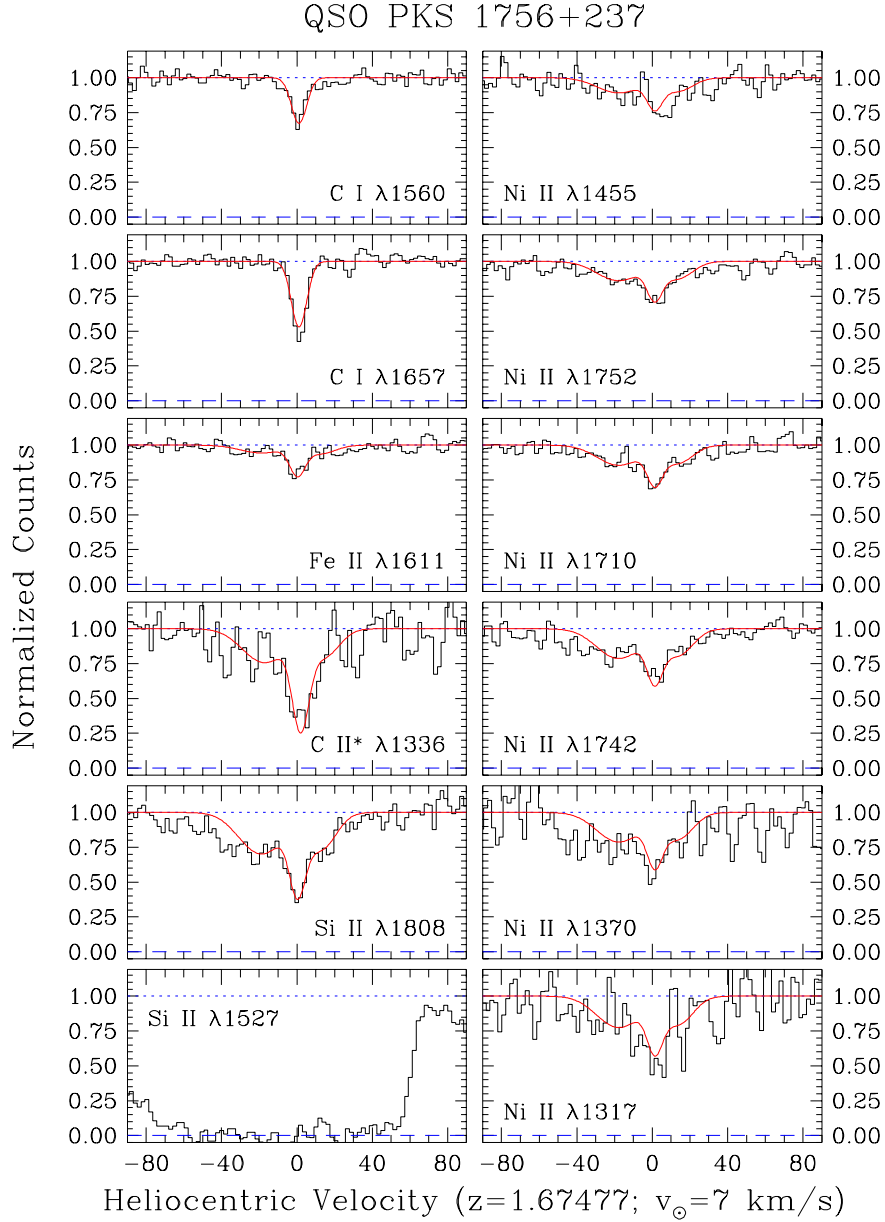}
\figcaption[figure1.ps]{Weak Ni II, Fe II, C I* and C I features along
with saturated Si II $\lambda$1527 in the $z=1.6748$ absorption system
toward PKS 1756$+$237. The spectra have been normalized and the
observed air wavelengths have been converted to vacuum heliocentric
velocities, with $v=0$ km s$^{-1}$ corresponding to $z=1.67477$.
Superimposed are Voigt model line profiles. \label{fig1}}
\end{figure}

PKS 1756$+$237 ($m_{\rm V}=18.0, z_{\rm em}=1.721$) was observed May 30
and 31, 1997 using the W. M. Keck Observatory I 10-m High Resolution
Echelle Spectrometer (HIRES, \cite{vog94}) for two hours.  The
signal-to-noise ratio varies from $\sim$ 5 at the blue end to $\gtrsim$
25 at longer wavelengths.  Two slightly offset grating tilt angles
yielded complete spectral coverage from 3190 -- 5055 \AA\ with a
resolution of 35,000 (FWHM $=$ 8.5 km s$^{-1}$).  Full details of the
spectra extraction and reduction procedures, along with a complete
analysis of all the observed spectral features, will be presented in
our companion paper (\cite{rot99}).  Among the surprises found in the
$z=1.6748$ absorption system are several Ni II lines, Fe II
$\lambda$1611, and C I $\lambda\lambda$1657,1560.  None of these were
expected to be present because of the low H I column density in this
absorber, log \NHI\ $\sim$ 19.5 (\cite{tur79}).  Figure \ref{fig1}
shows these lines plotted versus heliocentric corrected rest vacuum
velocity along with C II* $\lambda 1336$ and Si II $\lambda$1527.  We
have examined the entire spectrum for other absorption systems which
may confuse individual features in the $z=1.6748$ absorber.  The only
possibility is a system $z=1.7337$, probably associated with the QSO,
which could be contaminating our C I $\lambda$1560 line profile with Si
II $\lambda$1527 absorption.  However, we discount it as insignificant
because this absorber has a high ionization state and the corresponding
C II $\lambda$1335 line, expected to be much stronger than Si II
$\lambda$1527, is quite weak ($\approx$ 20 m\AA).

In this Letter we largely restrict our analysis to the weak features in
Figure \ref{fig1} which trace the highest column density neutral cloud
harboring C I.  Equivalent widths ($W_\lambda$) are measured by direct
integration across the line profiles.  Statistical $W_\lambda$ errors
are estimated from the signal-to-noise ratio in the adjacent continuum
(\cite{jen73}).  To account for continuum placement uncertainties we
inflate the errors $\sigma(W_\lambda)$ to equal 150\% of the formal
statistical values.  The resulting rest equivalent width measurements
and column densities resulting from a linear curve-of-growth ($N_{\rm
linear}$) are given in Table \ref{tab1}.  The good $N_{\rm linear}$
agreement among different absorption lines from the same species
suggests that most of these weak features are unsaturated.  Therefore
our column density measurements will be largely insensitive to the
adopted model line profile.  Our three-component Voigt profile
absorption model is given in Table \ref{tab2}.  We have convolved this
model with a Gaussian instrumental $b$-value of 5.1 km s$^{-1}$ to
produce the superimposed profiles presented in Figure \ref{fig1}.  The
total model column density in all three velocity components is given in
the last column of Table \ref{tab1}.  Note that the C I absorption,
which presumably arises from the densest material, is present only in
the central narrow velocity component.

\section{Abundances and Depletions}

Fe II, Ni II, and Si II are dominant ionization stages in neutral H I
material and thus measure gas-phase abundances.  Fe and Ni are both
Fe-peak elements and deplete readily onto dust grains in the Milky Way
ISM.  Si is formed via the $\alpha$-process and is much less
susceptible to dust, but its abundance relative to Fe and Ni may depend
on star formation history.  For these reasons, the interpretation of
relative abundance ratios is complicated (eg.\ \cite{vla98},
\cite{pet97}, \cite{lu96}).  The cosmic abundances for Si, Fe and Ni
have relative values of $1.00 : 0.91 : 0.050$ (\cite{and89}) while our
observed relative abundances (Table \ref{tab1}) are $1.00 : 1.16 :
0.031$.  The observed $2\,\sigma$ enhancement of Fe over Si is probably
not significant since the Si II $\lambda$1808 feature is 75\% deep and
therefore possibly saturated.  However, the observed deficit of Ni with
respect to Fe is more likely to be real.  The Fe abundance arises from
the very weak Fe II $\lambda$1611 feature which is unaffected by
saturation.  Therefore the Fe II column density uncertainty is simply
the 14\% equivalent width measurement error.  We estimate the Ni II
column density $1\,\sigma$ uncertainty to be $\pm$ 3.6\% from a
weighted average of the six linear curve-of-growth $1\,\sigma$ values.
This leads to an observed Ni/Fe abundance ratio of $0.027\pm0.004$,
significantly below the cosmic Ni/Fe value of 0.055.  Since both Fe and
Ni are Fe-peak elements, this observed deviation in the relative
gas-phase abundances from the cosmic value is likely due to depletion
onto grains.  Diffuse clouds in the Milky Way ISM come in roughly three
categories with different physical conditions and dust depletion
properties.  In order of increasing temperature and decreasing density,
these are cool disk clouds, warm disk clouds and halo clouds with
gas-phase Ni/Fe values of 0.012 -- 0.035, 0.028 -- 0.035, and 0.026 --
0.046, respectively (\cite{sem96}).  Our measured Ni/Fe abundance ratio
is completely consistent with galactic dust depletion, perhaps
suggestive of warm disk gas.  Unfortunately, our data do not extend to
long enough wavelengths to cover Zn II and Cr II. These lines should be
easily detectable, and the relative abundance ratios Ni/Zn, Fe/Zn or
Cr/Zn show a more pronounced variation with ISM environment because Zn
is observed to deplete very little, if at all, onto dust grains
(eg.\ \cite{pet97b}, \cite{rot95}, \cite{sem95}).

\section{Carbon I Excitation}

\begin{figure}
\epsscale{0.75}
\plotone{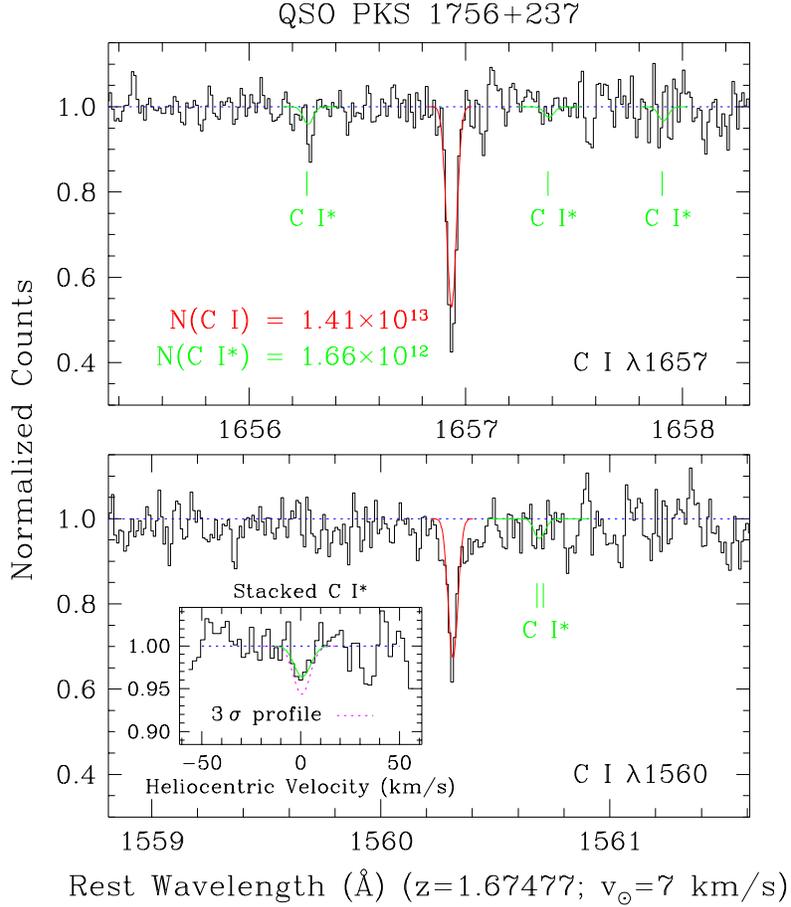}
\figcaption[figure2.ps]{C I $\lambda$1657 (top panel) and $\lambda$1560
(bottom panel) ground-state lines at $z=1.67477$ toward PKS
1756$+$237.  Superimposed on each line is our adopted model profile.
Also plotted are the expected strengths for the four excited-state C I*
lines with the redshifted CMBR as the only source of C I excitation.
In the inset figure (bottom panel) the four C I* lines have been
shifted to a common heliocentric velocity scale and averaged.  Similarly, the
expected C I* profile and our $3\,\sigma$ upper-limit have been stacked
and superimposed. \label{fig2}}
\end{figure}

We report the very surprising detection of C I ground-state absorption
in the $z=1.6748$ gas cloud (Figure \ref{fig2}, Table \ref{tab1}).
Since the lines are somewhat saturated, we have applied a $b=2.5$ km/s
Gaussian curve-of-growth to our equivalent width measurement errors to
estimate the uncertainty in our model column density value (Table
\ref{tab2}), yielding $N({\rm C\,I})=1.41\,(+0.30,\,-0.25) \times
10^{13}$ cm$^{-2}$.  We do not detect any C I* excited-state lines.
While in the Galactic ISM the dominant C I excitation mechanism is
neutral particle collisions, at high redshift the CMBR will make a
non-negligible contribution to the population of the $J=1$ level.  The
amount of pumping is given by the Boltzmann equation:  
\begin{equation}
T_{\rm CMBR}(1+z) = \frac{h\nu_{0\,\rightarrow1}}{k_{\rm B}}\left\{ \ln
\left( \frac{3\,n_{J=0}}{n_{J=1}}\right ) \right \}^{-1} \label{eqn1}
\end{equation} 
where $\nu_{0\,\rightarrow1}$ is the frequency of the
energy separation between the ground and first-excited states,
$(h\nu_{0\,\rightarrow1}/k_{\rm B})=23.595$ K (\cite{nus79}), and the
present-day CMBR temperature $T_{\rm CMBR}=2.726$ K (\cite{mat94}).
Replacing the number density ratio $n_{J=1}/n_{J=0}$  by the observable
quantity $N_{J=1}/N_{J=0}$ in Equation \ref{eqn1}, we predict an
excited-state column density of 
$N({\rm C\,I^*})=1.67\,(+0.35,\,-0.30)\times 10^{12}$ cm$^{-2}$.  On
Figure \ref{fig2} we superimpose the predicted C I* absorption
profiles.  As can be seen, the detection limit of our data is just
above the C I* line strengths produced by the redshifted $T=7.291$ K
CMBR.

To improve our C I* detection limit, we have shifted the observed data
points to a common velocity scale for each of the four excited-state
features (for the C I* $\lambda1561$ doublet we use an $f\,$-value
weighted average line center).  This average C I* spectrum is shown in
the inset to Figure \ref{fig2}, where we have superimposed the
predicted C I* profile produced using the same stacking method.  The
rms error of the inset stacked spectral region (S/N $=$ 45) corresponds
to an equivalent width error over nine pixels of 0.75 m\AA.  We inflate
this $\sigma_{W_\lambda}$ by 50\% for the same reason as before and
adopt a $1\,\sigma$ uncertainty in the stacked C I* equivalent width of
1.1 m\AA.  The stacked C I* profile has an equivalent width of 2.1
m\AA, which is consistent with no local excitation of the C I
fine-structure levels above CMBR pumping at the $2\,\sigma$ level.
However, since inspection of the stacked C I* spectrum suggests that
some absorption might be present, although at somewhat less than our
$2\,\sigma$ level, we have elected to adopt a 3$\,\sigma$ detection
upper-limit for this stacked C I* feature of $W_\lambda({\rm
C\,I^*})\leq 3.3$ m\AA, which implies a 3$\,\sigma$ C I* column density
upper-limit of $N({\rm C\,I^*})\leq 2.67\times 10^{12}$ cm$^{-2}$.
This leads to a 3$\,\sigma$ upper-limit on the C I excitation
temperature in the $z=1.6748$ absorber toward PKS 1756$+$237 of $T_{\rm
ex}\leq 8.54\,(+0.65,\,-0.56)$ K.  The stacked profile for this
$3\,\sigma$ detection limit is shown by the dashed line in the inset to
Figure \ref{fig2}.

In addition to measuring the CMBR temperature, our C I fine-structure
population measurement allows us to determine the physical conditions
within the absorbing gas cloud.  For collisional excitation by hydrogen
atoms, we employ the rate coefficients reported by \cite{lau77}.  For
UV pumping, we solve the detailed balance population equations for the
200 strongest C I UV lines (\cite{deb74}, \cite{jen83}, \cite{mor91},
\cite{zsa97}) and measure the UV field in units of the WJ1 average
interstellar radiation field (\cite{mor75} [Table 9], \cite{deb73},
\cite{wit73}).  In the absence of CMBR pumping and collisional
excitation, we find a 3$\,\sigma$ upper-limit on the UV field within
the $z=1.6748$ absorber of $\leq 14.3\,(+3.2,\,-2.6)$ times the
Galactic UV radiation field.  Alternatively, if collisions alone
populate the $J=1$ level and the gas kinetic temperature is 10 K,
typical of cool disk clouds, the hydrogen atom density is (3$\,\sigma$
upper-limit) $n_{\rm H}^{10{\rm K}}\leq 950\,(+1079,\,-369)$
cm$^{-3}$.  The C I collisional excitation rate is very sensitive at
low temperatures.  For example, at a slightly higher temperature (15 K)
the density limits are reduced significantly ($n_{\rm H}^{15{\rm
K}}\leq 191\,(+64,\,-44)$ cm$^{-3}$).  At kinetic temperatures similar
to less dense clouds or halo clouds (100 or 1000 K) the upper-limits
are further decreased to $n_{\rm H}^{100{\rm K}}\leq
25.1\,(+5.8,\,-4.6)$ and $n_{\rm H}^{1000{\rm K}}\leq
10.6\,(+2.4,\,-1.9)$ cm$^{-3}$.

These limits become more interesting when one includes the considerable
contribution to C I* excitation by the high redshift CMBR.  As a
result, observations of C I fine-structure lines in QSO absorbers can
yield far more sensitive probes of the physical conditions within
distant gas clouds than similar observations within the Milky Way can
provide.  For example, if we assume the CMBR temperature at $z=1.6748$
is given by the predicted big bang value of 7.291 K, we find that 
our UV field and hydrogen atom density upper-limits are substantially
lower.  With CMBR pumping, the ambient UV field must be less than
(3$\,\sigma$) $5.6\,(+3.3,\,-2.6)$ Galactic fields, $n_{\rm H}^{10{\rm
K}}\leq 371\,(+652,\,-224)$ cm$^{-3}$, $n_{\rm H}^{15{\rm K}}\leq
75\,(+54,\,-37)$ cm$^{-3}$, $n_{\rm H}^{100{\rm K}}\leq
9.8\,(+5.8,\,-4.6)$ cm$^{-3}$, and $n_{\rm H}^{1000{\rm K}}\leq
4.1\,(+2.4,\,-1.9)$ cm$^{-3}$.

In a survey of 27 Galactic stars using the {\it Copernicus}\/
satellite, \cite{jen83} did not report an excitation temperature lower
than 40 K and excited-state lines were detected in 23 lines of sight.
However, excitation temperatures were not determinable for nearly 25\%
of their sample, and all four of the sightlines lacking C I* absorption
had $N({\rm C\,I})$ values that were comparable to or below that found
in the $z=1.6748$ absorber.  Because of the redshifted CMBR, our limits
on the physical conditions within the $z=1.6748$ absorber are 2--3
times lower than the values in these four Milky Way sightlines.  It is
perhaps too early to tell whether high-redshift C I absorbers are
similar to Milky Way low column density sightlines, or if there is a
deficit of high column density C I absorbers toward QSOs.  The
discovery of more high-redshift C I QSO systems will eventually answer
the latter question, while higher quality C I spectra of the Milky Way
ISM are required to address the former.

\section{A Damped Ly$\alpha$ Absorber?}

\begin{figure}
\plotone{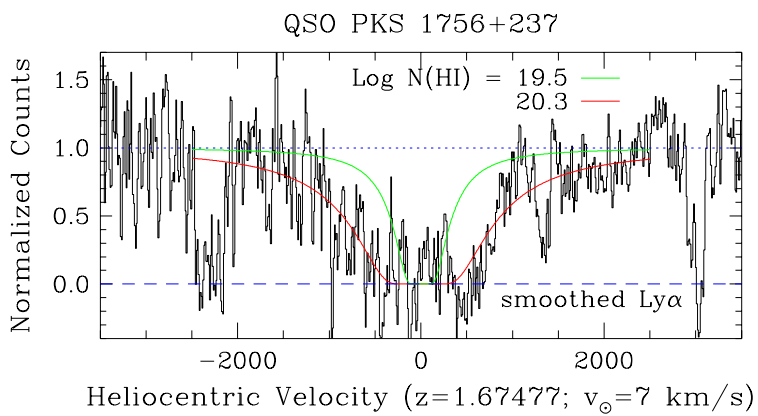}
\figcaption[figure3.ps]{Ly$\alpha$ absorption at $z=1.67477$ toward PKS
1756$+$237.  The data have been smoothed over 11 pixels and rebinned to
a dispersion of 10 km/s. Superimposed on the bottom panel are $N({\rm
H\,I})=3.16\times10^{19}$ and $2.0\times10^{20}$ cm$^{-2}$ Voigt
profiles with the same $b$-value and velocity found from the C I
lines.  \label{fig3}} 
\end{figure}

The presence of weak, singly-ionized metal lines and C I argue strongly
that the $z=1.6748$ absorber toward PKS 1756$+$237 has been erroneously
classified as a moderately low column density system and we have
assumed it is in fact damped.  Our spectral data extend far enough into
the blue to cover the H I absorption feature, and we have attempted to
verify the validity of this claim.  The Ly$\alpha$ line, smoothed and
rebinned to 20\% the resolution, is presented in Figure \ref{fig3} with
model Voigt profiles for the originally reported $\log N({\rm
H\,I})\sim19.5$ (\cite{tur79}) and a damped Ly$\alpha$ absorber
superimposed.  The data are noisy, the Ly$\alpha$ absorption is located
near the blue wing of the QSO Ly$\alpha$ emission ($z_{\rm em}=1.721$),
and it extends across two echelle orders, all of which complicate the
continuum placement.  For these reasons, we do not claim a conclusive H
I column density measurement.  However, we believe there is sufficient
evidence to reclassify this system as a damped Ly$\alpha$ absorber with
$\approx$ 6--10 times more neutral material than previously reported.

\acknowledgments

The authors wish to thank the staff at WMKO for their excellent
technical assistance in executing the observations.  KCR is grateful to
Antoinette Songaila for her advice and support as faculty sponsor at
the Institute for Astronomy, and to Ken Sembach for extremely helpful 
discussions and suggestions.  JMB acknowledges and thanks Alan Stockton
for his role as faculty contact during his AST699 Graduate Student
Research Project.  Funding for this work was provided in part by NASA
through Hubble Fellowship grant \#HF-01076.01-94A and grants
\#GO-05887.02-94A, \#GO-05888.01-94A awarded by the Space Telescope
Science Institute, which is operated by the Association of Universities
for Research in Astronomy, Inc., for NASA under contract NAS5-26555.

\begin{deluxetable}{lcccccc}
\tablecaption{Rest Equivalent Widths and Column Densities \label{tab1}}
\tablehead{
\colhead{Transition} & \colhead{$\lambda$} & \colhead{$f$\tablenotemark{a}} & \colhead{S/N} & \colhead{$W_\lambda \pm 1\sigma$} & 
\colhead{$N\tablenotemark{b}_{\rm linear}\pm 1\sigma$} &
\colhead{$N\tablenotemark{c}_{\rm model}$}\nl
\colhead{~} & \colhead{(\AA)} & \colhead{~} & \colhead{~} & \colhead{(m\AA)} & \colhead{(cm$^{-2}$)} & \colhead{(cm$^{-2}$)}
} 
\startdata
Fe II $\lambda$1611 & 1611.2005 & 0.00102 & 26 & $27.9\pm3.9$ & $(1.19\pm0.17)\times10^{15}$ & $1.23\times10^{15}$\nl
Si II $\lambda$1808 & 1808.0126 & 0.005527& 19 & $153.5\pm7.6$&
$(9.60\pm0.48)\times10^{14}$ & $1.06\times10^{15}$\nl
Ni II $\lambda$1455 & 1454.8420 & 0.05954 & 17 & $44.6\pm5.5$ &
$(4.00\pm0.49)\times10^{13}$ & $3.30\times10^{13}$\nl
Ni II $\lambda$1317 & 1317.2170 & 0.1458  &  6 & $83\pm15$    &
$(3.71\pm0.67)\times10^{13}$ & $''$               \nl
Ni II $\lambda$1370 & 1370.1320 & 0.1309  &  9 & $92\pm12$    &
$(4.23\pm0.55)\times10^{13}$ & $''$               \nl
Ni II $\lambda$1710 & 1709.6000 & 0.06884 & 25 & $58.7\pm4.7$ & $(3.30\pm0.26)\times10^{13}$ & $''$               \nl
Ni II $\lambda$1742 & 1741.5490 & 0.1035  & 24 & $84.5\pm5.1$ & $(3.04\pm0.18)\times10^{13}$ & $''$\nl
Ni II $\lambda$1752 & 1751.9100 & 0.06380 & 26 & $65.5\pm4.8$ & $(3.78\pm0.28)\times10^{13}$ & $''$\nl
C I $\lambda$1657   & 1656.9283 & 0.1405  & 25 & $29.2\pm2.4$ & $(0.855\pm0.070)\times10^{13}$ & $1.41\times10^{13}$\nl
C I $\lambda$1560   & 1560.3092 & 0.08041 & 23 & $22.1\pm3.1$ & $(1.28\pm0.18)\times10^{13}$ & $''$\nl
C II$^*$ $\lambda$1336&1335.7077& 0.1149  &  7 & $92\pm12$    &
$(5.07\pm0.66)\times10^{13}$ & $1.10\times10^{14}$
\enddata

\tablenotetext{a}{Oscillator strengths ($f$) from \cite{mor91} except for
FeII $\lambda$1611 $f$-value taken from \cite{car95}}
\tablenotetext{b}{Column densities assuming linear curve-of-growth, $N_{\rm linear}=(1.1296\times 10^{17})W_\lambda \lambda^{-2} f^{-1}$ cm$^{-2}$}
\tablenotetext{c}{Voigt profile model column densities:  Fe II and Ni II
values result from a 3-cloud sum, C I value includes only a single velocity 
component}

\end{deluxetable}

\begin{deluxetable}{lccc}
\tablewidth{0pt}
\tablecaption{Three-Component Voigt Absorption Model \label{tab2}}
\tablehead{
\colhead{Ion} & \colhead{log($N_1$)\tablenotemark{a}} & 
\colhead{log($N_2$)\tablenotemark{b}} & 
\colhead{log($N_3$)\tablenotemark{c}}} 
\startdata
C I   &$\cdots$& 13.15 &$\cdots$\nl
Fe II & 14.58  & 14.75 & 14.47  \nl
Ni II & 13.15  & 13.00 & 12.95  \nl
Si II & 14.58  & 14.65 & 14.37  \nl
C II* & 13.29  & 13.90 & 13.05
\enddata
\tablenotetext{a}{$v_1=-19$ km/s, $b_1=15$ km/s}
\tablenotetext{b}{$v_2=0$ km/s, $b_2=2.5$ km/s}
\tablenotetext{c}{$v_3=+12$ km/s, $b_3=10$ km/s}
\end{deluxetable}

\clearpage

\end{document}